\documentclass[aps,prl,reprint]{revtex4-1}
\usepackage{graphicx}

\begin{document}
\title{The co-existence of states in $p53$ dynamics driven by $miRNA$}
\author{Md. Jahoor Alam$^1$}
\author{Shazia Kunvar$^2$}
\author{R.K. Brojen Singh$^{1}$}
\email{brojen@mail.jnu.ac.in}
\affiliation{$^1$School of Computational and Integrative Sciences, Jawaharlal Nehru University, New Delhi-110067, India \\
$^2$ Department of Bioinformatics, Banasthali Vidyapith,Rajasthan-304022, India.}

\begin{abstract}
The regulating mechanism of $miRNA$ on $p53$ dynamics in $p53-MDM2-miRNA$ model network incorporating reactive oxygen species ($ROS$) is studied. The study shows that $miRNA$ drives $p53$ dynamics at various states, namely, stabilized states and oscillating states (damped and sustain oscillation). We found the co-existence of these states within certain range of the concentartion level of $miRNA$ in the system. This co-existence in $p53$ dynamics is the signature of the system's survival at various states, normal, activated and apoptosis driven by a constant concentration of $miRNA$.
\end{abstract}

\maketitle

$Introduction.-$ The p53, tumor suppressor protein, attracted the interest of researchers because of its important role in preventing cell to become cancer \cite{lan,whi}. It acts as a key regulator in the cellular network and response to a variety of cellular stress, including DNA damage, hypoxia, nucleotide depletion, nitric oxide and aberrant proliferative signals (such as oncogene activation) \cite{lan,mol}. But in most cases of human cancer cell, p53 tumor suppressor signaling pathway usually found in inactivated condition \cite{lan}. Its activation results in the fulfillment of key cellular processes, for example, cell-cycle arrest, senescence and most importantly tumor clearance to prevent cancer cell formation \cite{zam}.  Further, activated p53 protein safeguards the organism against the propagation of cells that carry damaged DNA with potentially oncogenic mutations \cite{mol}. It has been reported that activation of p53 functions via the inhibition of MDM2 protein can be regarded as an effective approach in cancer therapy \cite{mic}. Because MDM2 acts as a negative feedback regulator (inhibitor) to p53 by binding itself to p53, and then physically blocking its ability to transactivate gene expression, and stimulating its degradation \cite{wan,kub,fan}. Further, the interaction of N-terminal domain of MDM2 with transactivation domain of p53 (p53TAD) performs a significant role in the regulation of the G1 checkpoint of the cell cycle and cell function\cite{boy,che}. 

ROS (Reactive oxygen species) are chemically reactive molecules containing oxygen ions and peroxides\cite{ame}. They are synthesized from normal metabolism of oxygen as a natural byproduct and play important roles in cell signaling and homeostasis \cite{dev,row}. However, ROS level inside cell can be elevated by UV irradiation or heat exposure which can drive the cell at different stress states \cite{dev}. High level of ROS can promote DNA damage, and may probably lead the cell to mutagenesis, carcinogenesis and aging \cite{ame,row,proc}. However, the role of ROS in driving the cell at different states, namely, normal, stress, cancerous and apoptosis is still not fully studied.

MicroRNAs (miRNAs) are small noncoding RNA molecule of size 20-24 nucleotides, and are powerful regulators of transcriptional and post transcriptional gene expression which regulate both physiological and pathological processes such as cellular development and cancer \cite{tho,lee1,wil}. miR-125b is a brain-enriched miRNA which acts as a negative regulator of p53 both in zebrafish and human \cite{le,zha1,win}. Overexpression of miR-125b suppresses the endogenous level of p53 protein and represses to apoptosis in human neuroblastoma cells and human lung fibroblast cells \cite{le}. Decrease in level of miR-125b leads to enhance the level of p53 and induces apoptosis in human neuroblastoma and human lung fibroblast cells \cite{le,zha1}. However, the regulating mechanism of miR-125b with p53 is not fully studied. The dynamics of p53 and its response to the miR-125b regulation are still open questions. In the present study, we try to answer some of these fundamental questions based on basic model built from available experimental reports. 

$p53-MDM2-miRNA~model.-$ The model we consider (Fig. 1) is integration of p53-Mdm2 regulatory network \cite{pro} with stress inducers ROS via DNA damage \cite{row} and miRNA which interact with $p53\_MDM2$ \cite{win}. In this model we asuume that $miRNA$s are supposed to be constantly produced in the nucleus either from their own genes or encode from introns (non-coding sequence) with a rate $k_1$ \cite{win}. ROS synthesis is assumed to occur with a rate of $k_{14}$. This ROS synthesis triggers DNA damage with a rate of $k_{16}$ \cite{row}. Then this DNA damage leads to the activation of $ARF$ with a rate $k_{18}$ \cite{lee} followed by the degradation of $ARF$ with a rate of $k_{17}$. Further, the activated $ARF$ protein binds to $MDM2$ with a rate of $k_{20}$ to control ubiquitination of $p53$ \cite{zha}. The $ARF$ and $MDM2$ interaction results into to the formation of $ARF\_MDM2$ complex \cite{kha}. The formation of $ARF\_MDM2$ complex reduces the concentration level of $MDM2$ in the systems which in turn alters the behaviour of $p53$ \cite{kha}. On the other hand, dissociation of $ARF\_MDM2$ complex with a rate $k_{21}$ helps the degradation of $MDM2$ population and recruit activated ARF. $miRNA$ directly interacts with $p53\_mRNA$ to form $miRNA\_p53\_mRNA$ complex at a rate $k_2$ \cite{wil}. The ubiquitination of $p53\_mRNA$ is done via $miRNA$ which occurs with a rate $k_3$ \cite{win}. The synthesis of $p53$ takes place through transcription of $p53\_mRNA$ with a rate $k_4$. Further, this $p53$ synthesis depends on the available $p53\_mRNA$ concentration level. At normal condition $p53$ is generally bound to $MDM2$ with a rate $k_{10}$ recruiting a complex $p53\_MDM2$ and after which the dissociation of the complex ubiquitinates $p53$ with a rate $k_{11}$ and $MDM2$ with a rate $k_{12}$ \cite{gev,jah} exhibiting oscillatory behavior of $p53$ in the model. Further, $p53$ is found to be transcription factor which interact with $MDM2$ gene and leads to the production of $MDM2\_mRNA$ with a rate $k_7$ \cite{pro}. Hence, the $MDM2\_mRNA$ provides intermediary link between $p53$ and $MDM2$. The self ubiquitination of $MDM2\_mRNA$ is assumed to be with a rate $k_9$. $MDM2\_mRNA$ synthesize $MDM2$ protein with a rate $k_8$. The self ubiquitination of $MDM2$ is assumed to be with a rate $k_{13}$. The molecular species involved in this model are listed in Table 1 (Supplementary file), and the biochemical reaction channels involved in the model network with their descriptions, kinetic laws and values of the rate constants used in our simulations are given in Table 2 (Supplementary file).
\begin{figure}
\label{}
\includegraphics[height=330 pt,width=9cm]{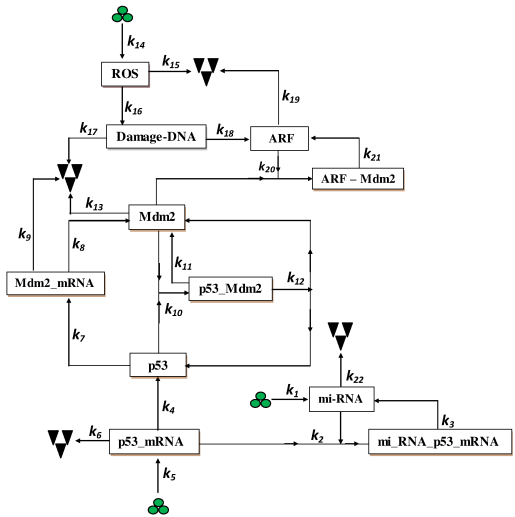}
\caption{A schematic diagram of p53-Mdm2-mi-RNA network model.} 
\end{figure}

Consider the state of the system be described by a state vector given by, ${\bf x}(t)$=$\left[x_1(t), x_2(t),\dots, x_N(t)\right]^T$, where, $\{x\}$ is the set of concentrations of the respective molecular species, $N=11$ and $T$ is the transpose of the vector. The model biochemical network (Fig. 1) described by the twenty two reaction channels (Table 2) can be described by the following coupled ordinary differential equations (ODE) using Mass action law of chemical kinetics,
\begin{eqnarray}
\label{nl}
\frac{dx_i(t)}{dt}=F_i\left[x_1(t),x_2(t),\dots,x_N(t)\right]
\end{eqnarray}
where, $i=1,2,...,N$ and $F_i$ is the ith function whose form is given in Supplementary file. The non-linear coupled $N$ ODEs (\ref{nl}) (Supplementary fine) of $p53-MDM2-miRNA$ model are solved using 4th order Runge-Kutta method which is the standard algorithm for numerical integration \cite{pre} to find the dynamics of the system variables. The simulation is done for 10 days using the parameter values given in Supplementary file (Table 2) and starting from an initial condition.

$ROS~driven~p53~phase~transition.-$
The concentration of $ROS$ in the system drives the system dynamics at different states which may correspond to various temporal cellular states. The simulation is done first keeping $k_{miRNA}=0$ throughout the numerical experiment, and changing the parameter $k_{ROS}$ (Fig. 2). Since $k_{ROS}$ is the rate of creation of $ROS$, the concentration of $ROS$ synthesized in the system is proportional to $k_{ROS}$. The $p53$ level in the system is maintained at stabilized state with minimum concentration level for sufficiently small values of $k_{ROS}$ ($k_{ROS}\le 0.00002$) which may correspond to normal state of the system. As the value of $k_{ROS}$ increases slightly ($0.0002\le k_{ROS}\langle 0.002$) the dynamics cross over from stable state to damped oscillation state (Fig. 2 B) where the dynamics preserves stable condition for certain interval of time ([0-7] days), and then it becomes activated (for time$\ge$ 7 days) induced by $k_{ROS}$. This result suggests that as the concentration of $ROS$ increases in the system, it causes more DNA damage due to which $p53$ dynamics become stressed and exhibits an oscillatory pattern. Further increase in the value of $k_{ROS}$ ($0.002\le k_{ROS}\langle 0.008$) leads the $p53$ dynamics to damped oscillation for some interval of time then to sustained oscillation with increasing amplitude (Fig. 2 C and D; Fig. 3 upper left panel). The sustain oscillation indicates that the $p53$ is strongly activated (the stress is maximum). 
\begin{figure}
\label{}
\includegraphics[height=250 pt,width=8cm]{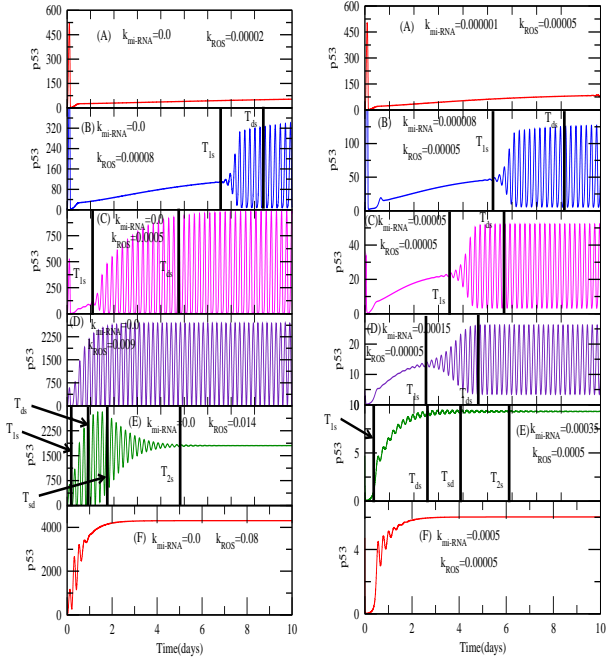}
\caption{(A)The p53 temporal behaviour when ROS act as a stress inducer whereas $mi-RNA$ creation rate kept fixed. (B)The p53 temporal behaviour when $mi-RNA$ act as a stress inducer whereas $ROS$ creation rate kept fixed.} 
\end{figure}

Now, excess increase in ROS concentration ($k_{ROS}\ge 0.015$) drives the $p53$ dynamics from sustain to damped oscillation (Fig. 2 E), after which $p53$ state is switched to stabilized state (Fig. 2 E and F; Fig. 3 upper left panel). This suggests that extreme values of $k_{ROS}$ may cause very high DNA damage, such that the damage could not able be repaired back, which could be the condition of apoptotic phase. 

$Role~of~miRNA~on~p53~dynamics.-$
The interaction of $miRNA$ with $p53$ is done via $p53\_mRNA$ complex in indirect fashion. The impact of $miRNA$ on $p53$ was studied by keeping fixed $k_{ROS}=0.00005$ and allowing to change the values of $k_{miRNA}$ (Fig. 2 right panels). Similarly, as obtained in $ROS$ case, we got three different states namely stable, damped with sustain oscillation and again stable state of $p53$ driven by $miRNA$ (Fig. 2 right panels). The small values of $k_{miRNA}$ ($k_{miRNA}\langle 0.000001$) could not able to provide significant stress to $p53$ dynamics, and maintains at stabilized state (Fig. 2 A). The further increase in $k_{miRNA}$ values ($0.00001\le k_{miRNA}\langle 0.0002$) the dynamics still maintains stability upto certain interval of time (Fig. 2 B, C, D), after which the dynamics is switched to damped oscillation (weakly activated) for short interval of time and then to sustain oscillation (strongly activated). Further increase in $k_{miRNA}$ compels the dynamics to stabilized state again with low concentration level (Fig. 2 F). This suggests that the increase in concentration of $miRNA$ in the system drives the system at various stress states, lowering $p53$ concentration level \cite{le,sel}. The excess $k_{miRNA}$ values induce lowering of $p53$ concentration level even below normal stabilized $p53$ state indicating the possibility of switching stress state to cancerous state \cite{sel}.
\begin{figure}
\label{}
\includegraphics[height=200 pt,width=8cm]{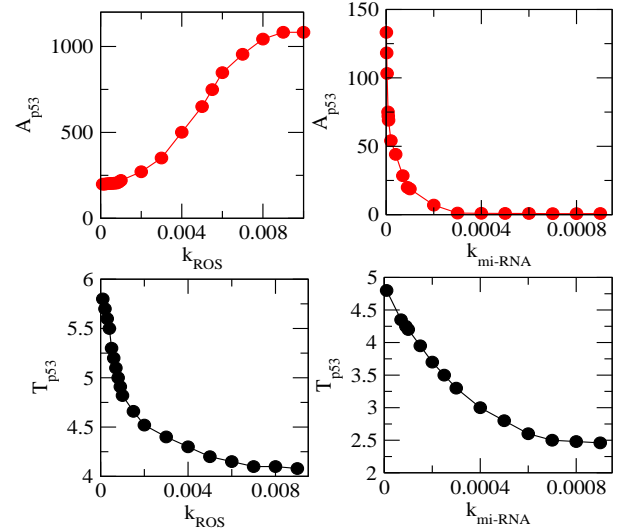}
\caption{A comparative plot for the amplitude verses $k_{ROS}$ in first panel and $k_{mi-RNA}$ in 2nd panel. Similarly, comparative plot for the time period variation verses $k_{ROS}$ in 3rd panel and $k_{mi-RNA}$ in 4th panel.} 
\end{figure}

$Co-existence~of~states.-$
The phase transition like behaviour of the system dynamics induced by ROS and miRNA concentrations available in the system can be well characterized by analysing the nature of transition time of the p53 dynamics. We define $T_{1s}$ to be the transition time below ($t\langle T_{1s}$) which the dynamics shows stable state (does not show any oscillation) and above which the dynamics shows oscillatory behaviour. We further define second transition time, $T_{ds}$ which separates increasing damped and sustain oscillations (Fig. 2). Similarly, $T_{sd}$ and $T_{2s}$ are taken as transition times separating sustain and damped oscillation, and damped oscillation and stabilized state. We then calculated $T_{1s}$, $T_{ds}$, $T_{sd}$ and $T_{2s}$ as a function of $k_{ROS}$ (Fig. 4 upper panel) where the regimes for $T\langle T_{1s}$ and $T\rangle T_{2s}$ corresponds to stabilized states, regimes between $T_{ds}\rangle T\rangle T_{1s}$ and $T_{2s}\rangle T\rangle T_{sd}$ corresponds to damped states and $T_{sd}\rangle T\rangle T_{ds}$ indicates the sustain oscillation state regime.
\begin{figure}
\label{}
\includegraphics[height=220 pt,width=8cm]{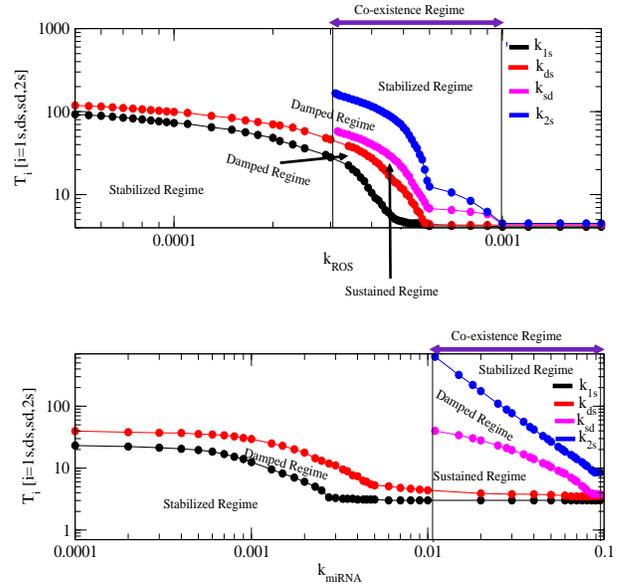}
\caption{A phase diagram showing impact of $k_{ROS}$ on stability of p53 as well as impact of $k_{mi-RNA}$ on stability of p53 protein.} 
\end{figure}

The results indicate that there is a certain range of $k_{ROS}$ (region bounded by two lines) where one can find the four states together including two stable states for any value of $k_{ROS}$ (Fig. 4 upper panel). This means that for any concentration of $ROS$ in the system corresponding to any values within this range, the $p53$ dynamics will stay stable for some interval of time, then it will start activated to reach maximum activation within certain interval of time and after sometime it will stay stable again. In the other regimes, at most we can find three states.

Similarly, the co-existence of the four states can be obtained in the case of $miRNA$ induced $p53$ dynamics also. Within this co-existence regime, the regions of damped, sustain and stabilized states are different as compared to $ROS$ induced $p53$ dynamics. This co-existence of the states indicate that exposure of the system to constant $miRNA$ concentration can drive the system from normal to stress and then to apoptosis. 

$Conclusion.-$
$p53$ is found to be a versatile protein which can interact with a number of protein and participate in many biologically important pathway. There are a number of factors which can induce cellular stress, such as environmental factors (UV, IR etc), stress inducing molecules ($ROS$, $miRNA$, nitric oxide and many other molecules). The variation in concentration of reactive oxygen species in cellular system leads to the changes in the $p53$ dynamics (various stress states) with overall enhancement in its concentration level in the cell. Further, the introduction $miRNA~125b$ to the system shows inhibitory effect on $p53$ production and switching of stress states by varying $miRNA~125b$ concentration \cite{le,zha1}. The obtained results are quite interesting and provide many hidden information regading the activity of $miRNA~125b$ that it can probably switch the system to cancerous state. Various experimental studies reported that concentration of $miRNA~125b$ increases in different cancer cell lines especially in breast cancer, leukemia and uterus cancer cell lines. Therefore, it is very important to study $miRNA$ in depth in order to understand other roles of it in regulating cancerous cells. 

Our study shows that significant activity of $miRNA$ can be seen only when the the system is slightly activated by $ROS$ but this process is not needed to study $ROS$ activity. This means that there is always a competition between $ROS$ and $miRNA$ which is needed to be investigated extensively. Moreover, the impact of the $miRNA$ on $p53$ regulatory pathway should be further studied in stochastic system in order to capture the state switching mechanism quantitatively and to understand the role of noise in the cellular process. 

$Acknowledgments.-$ This work is financially supported by Department of Science and Technology (DST), New Delhi, India under sanction no. SB/S2/HEP-034/2012.


\bibliographystyle{apsrev4-1} 
\bibliography{xampl} 

\end{document}


\title{The co-existence of states in $p53$ dynamics driven by $miRNA$}
\author{Md. Jahoor Alam$^a$}
\author{Shazia Kunvar$^b$}
\author{R.K. Brojen Singh$^{a}$}
\email{brojen@mail.jnu.ac.in}
\affiliation{$^a$School of Computational and Integrative Sciences, Jawaharlal Nehru University, New Delhi-110067, India \\
$^b$ Department of Bioinformatics, Banasthali Vidyapith,Rajasthan-304022, India.}

\maketitle

The model biochemical network (Fig. 1) described by the twenty two reaction channels (Table 2) can be described by the following coupled ordinary differential equations (ODE) using Mass action law of chemical kinetics,
\begin{eqnarray}
\label{ma}
\frac{dx_1}{dt}&=&k_4x_4-k_{10}x_1x_2+k_{12}x_5\\
\frac{dx_2}{dt}&=&k_8x_3-k_{10}x_1x_2+k_{11}x_5+k_{12}x_5
-k_{13}x_2-k_{20}x_2x_8\\
\frac{dx_3}{dt}&=&k_7x_1-k_9x_3\\
\frac{dx_4}{dt}&=&-k_2x_{10}x_4+k_5-k_6x_4\\
\frac{dx_5}{dt}&=&k_{10}x_2x_1-k_{11}x_5-k_{12}x_5\\
\frac{dx_6}{dt}&=&k_{14}-k_{15}x_6-k_{16}x_6\\
\frac{dx_7}{dt}&=&k_{16}x_6-k_{17}x_7\\
\frac{dx_8}{dt}&=&k_{18}x_7-k_{19}x_8-k_{20}x_8x_2+k_{21}x_9\\
\frac{dx_9}{dt}&=&k_{20}x_8x_2-k_{21}x_9\\
\frac{dx_{10}}{dt}&=&k_1-k_2x_{10}x_4+k_3x_{11}-k_{22}x_{10}\\
\frac{dx_{11}}{dt}&=&k_2x_{10}x_4-k_3x_{11}
\end{eqnarray}
The set of ODEs can be written in compact form as in the following,
\begin{eqnarray}
\label{ode}
\frac{d{\bf x}(t)}{dt}={\bf F}(x_1,x_2,\dots,x_N)
\end{eqnarray}
where, ${\bf F}=\left[F_1,F_2,\dots,F_N\right]^T$ is the functional vector.
The time evolution of the state vector ${\vec x}(t)$ can be obtained by numerically solving the non-linear coupled differential equations (1)-(11) using standard 4th order Runge-Kutta algorithm for numerical integration \cite{pre}. 

\subsection{Stability analysis}

The fixed or equilibrium points of the ODE given by equation (\ref{ode}) can be obtained by putting $\frac{d{\bf x}(t)}{dt}=0$ and solving for $x_1^*$, $x_2^*$, ..., $x_N^*$ from these equations. In our model described by mathematical equations (1)-(11), we have the following equilibrium points,
\begin{eqnarray}
x_1^*&=&\left[\frac{k_4k_5k_9}{k_7k_8k_{10}k_{11}}\left\{k_{13}+\frac{{\bf k_{ROS}}k_{16}k_{18}k_{20}}{k_{17}k_{18}(k_{15}+k_{16})}\right\}\frac{k_{11}+k_{12}}{k_6+\frac{k_2}{k_{22}}{\bf k_{miRNA}}}\right]^{1/2}\\
x_2^*&=&\left[\frac{k_4k_5k_7k_8(k_{11}+k_{12})}{k_9k_{10}k_{11}\left\{k_{13}+\frac{{\bf k_{ROS}}k_{16}k_{18}k_{20}}{k_{17}k_{18}(k_{15}+k_{16})}\right\}}\frac{1}{k_6+\frac{k_2}{k_{22}}{\bf k_{miRNA}}}\right]^{1/2}\\
x_3^*&=&\left[\frac{k_4k_5}{k_8k_{10}k_{11}}\left\{k_{13}+\frac{{\bf k_{ROS}}k_{16}k_{18}k_{20}}{k_{17}k_{18}(k_{15}+k_{16})}\right\}\frac{k_{11}+k_{12}}{k_6+\frac{k_2}{k_{22}}{\bf k_{miRNA}}}\right]^{1/2}\\
x_4^*&=&\frac{k_5k_{22}}{k_2+k_1k_{6}}\\
x_5^*&=&\frac{k_4k_5k_{22}}{k_{11}(k_2+k_1k_{6})}\\
x_6^*&=&\frac{{\bf k_{ROS}}}{k_{15}+k_{16}}\\
x_7^*&=&\frac{{\bf k_{ROS}}k_{16}}{k_{17}(k_{15}+k_{16})}\\
x_8^*&=&\frac{{\bf k_{ROS}}k_{16}k_{18}}{k_{17}k_{19}(k_{15}+k_{16})}\\
x_9^*&=&\left[\frac{k_4k_5k_7k_8(k_{11}+k_{12})}{k_9k_{10}k_{11}\left\{k_{13}+\frac{{\bf k_{ROS}}k_{16}k_{18}k_{20}}{k_{17}k_{18}(k_{15}+k_{16})}\right\}}\frac{1}{k_6+\frac{k_2}{k_{22}}{\bf k_{miRNA}}}\right]^{1/2}\nonumber\\
&&\times\frac{{\bf k_{ROS}}k_{16}k_{18}k_{20}}{k_{17}k_{19}k_{21}(k_{15}+k_{16})}\\
x_{10}^*&=&\frac{k_1}{k_{22}}\\
x_{11}^*&=&\frac{k_1k_2k_5}{k_3(k_2+k_1k_{6})}
\end{eqnarray}
The stabilized state of p53 ($x_1^*$) and Mdm2 ($x_2^*$) are dependent on the values of the parameters $k_{miRNA}$ and $k_{ROS}$, and there is competition between these two parameters affecting stabilized states of p53 and Mdm2. Keeping $k_{miRNA}$ to a constant value, equation (13) shows that $x_1^*\propto \sqrt{1+Ak_{ROS}}$, where, $A=\frac{k_{16}k_{18}k_{20}}{k_{13}k_{17}k_{18}(k_{15}+k_{16})}$ which drives the low equilibrium state (may be normal state where $x_1^*$ is maintained minimum value) at low $k_{ROS}$ to the higher equilibrium state (may be apoptosis state where $x_1^*$ is maintained at high value) as $k_{ROS}$ increases. However, in the case of Mdm2, the scenario is opposite, where $x_2^*\propto \frac{1}{\sqrt{1+Ak_{ROS}}}$, and $k_{ROS}$ drives the higher Mdm2 equilibrium state to lower equilibrium state.
 
Further, if $k_{ROS}$ and other rates are kept constant, $x_1^*\propto\frac{1}{\sqrt{B+k_{miRNA}}}$ and $x_2^*\propto\frac{1}{\sqrt{B+k_{miRNA}}}$, where $B$ is a constant given by $B=\frac{k_6k_{22}}{k_2}$. This means that as $k_{miRNA}$ increases, $k_{miRNA}$ drives the higher equilibrium state of both p53 and Mdm2 to lower equilibrium states.

\newpage


\begin{table*}
\begin{center}
{\bf Table 1 - List of molecular species} 
\begin{tabular}{|l|p{4cm}|p{5cm}|p{1.5cm}|}
 \hline \multicolumn{4}{}{} \\ \hline

\bf{ S.No.}    &   \bf{Species Name}    &    \bf{Description}                    &  \bf{Notation}     \\ \hline
1.             &    $p53$               & Unbounded $p53$ protein                &  $x_1$ \\ \hline
2.             &    $Mdm2$              & Unbounded $Mdm2$ protein               &  $x_2$ \\ \hline
3.             &    $Mdm2\_mRNA$        & $Mdm2$ messenger $mRNA$                &  $x_3$  \\ \hline
4.             &    $p53\_mRNA$         & $p53$ messenger $mRNA$                 &  $x_4$  \\ \hline
5.             &    $Mdm2\_p53$         & $Mdm2$ with $p53$ complex              &  $x_5$  \\ \hline
6.             &    $ROS$               & Reactive Oxygen Species                &  $x_6$  \\ \hline
7.             &    $Dam\_DNA$           & Damage DNA                             &  $x_7$  \\ \hline
8.             &    $ARF$               & Alternative Reading Frame protein      &  $x_8$  \\ \hline
9.             &    $ARF\_Mdm2$         & $ARF$ and $Mdm2$ complex               &  $x_9$  \\ \hline
10.            &    $mi-RNA-125b$       & Micro RNA 125b                         &  $x_{10}$  \\ \hline
11.            &    $mi-RNA\_p53-mRNA$  & Micro RNA 125b and $p53\_mRNA$ complex &  $x_{11}$  \\ \hline
\end{tabular}
\end{center}
\end{table*}

\newpage


\begin{table*}
\begin{center}
{\bf Table 2 List of chemical reaction, Kinetic Laws and their rate constant} 
\begin{tabular}{|l|p{2.5cm}|p{4cm}|p{2.5cm}|p{3cm}|p{2cm}|}
\hline \multicolumn{6}{}{}\\ \hline

${\bf S.No}$ & ${\bf Reaction}$ & ${\bf Name~of~the~process}$ & ${\bf Kinetic~Law}$ & ${\bf Rate~Constant}$ & ${\bf References}$ \\ \hline
1 & $\phi\stackrel{k_{1}}{\longrightarrow}x_{10}$ & Micro RNA creation & $k_2$ & $1\times 10^{-4}{sec}^{-1}$ & \cite{le,lee1,win}\\ \hline
2 & $x_{10}+x_4\stackrel{k_{2}}{\longrightarrow}x_{11}$ & Synthesis of miRNA and p53\_mRNA complex & $k_2 \langle x_{10}\rangle\langle x_4\rangle $ & $2\times 10^{-2}{sec}^{-1}$ & \cite{le,lee1}\\ \hline
3 & $x_{11}\stackrel{k_{3}}{\longrightarrow}x_{10}$ & $miRNA\_p53\_mRNA$ degradation & $k_3 \langle x_{11}\rangle$ & $1\times 10^{-4}{sec}^{-1}$ & \cite{le,lee1} \\ \hline
4 & $x_4\stackrel{k_{4}}{\longrightarrow}x_1+x_4$ & p53 mRNA translation & $k_4 \langle x_4\rangle$ & $8\times 10^{-2}{sec}^{-1}$ & \cite{mol,pro} \\ \hline
5 & $\phi\stackrel{k_{5}}{\longrightarrow}x_4$ & $p53\_mRNA$ synthesis & $k_5$  & $1\times 10^{-3}{sec}^{-1}$ & \cite{mol,pro}\\ \hline
6 & $x_4\stackrel{k_{6}}{\longrightarrow}\phi$ & $p53\_mRNA$ degradation & $k_6 \langle x_4\rangle$ & $1\times 10^{-4}{sec}^{-1}$ & \cite{mol,pro} \\ \hline
7 & $x_1\stackrel{k_{7}}{\longrightarrow}x_1+x_3$ & $Mdm2\_mRNA$ synthesis & $k_7 \langle x_1\rangle$ & $1\times 10^{-4}{sec}^{-1}$ & \cite{mol,pro,jah} \\ \hline
8 & $x_3\stackrel{k_{8}}{\longrightarrow}x_2+x_3$ & $Mdm2$ synthesis & $k_8 \langle x_3\rangle$ & $495\times 10^{-5}{sec}^{-1}$ & \cite{mol,pro,jah} \\ \hline
9 & $x_3\stackrel{k_{9}}{\longrightarrow}\phi$ & $Mdm2\_mRNA$ degradation & $k_9 \langle x_3\rangle$ & $1\times 10^{-4}{sec}^{-1}$ & \cite{mol,pro,jah} \\ \hline
10 & $x_1+x_2\stackrel{k_{10}}{\longrightarrow}x_5$ & $p53\_Mdm2$ complex formation & $k_{10} \langle x_1\rangle \langle x_2\rangle$ & $1155\times 10^{-3}{sec}^{-1}$ & \cite{mol,pro,jah} \\ \hline
11 & $x_5\stackrel{k_{11}}{\longrightarrow}x_2$ & Mdm2 creation & $k_{11} \langle x_5\rangle$ & $825\times 10^{-4}{sec}^{-1}$ & \cite{mol,pro,jah} \\ \hline
12 & $x_5\stackrel{k_{12}}{\longrightarrow}x_1+x_2$ & Dissociation of $p53\_Mdm2$ complex & $k_{12} \langle x_5\rangle$ & $1155\times 10^{-5}{sec}^{-1}$ & \cite{mol,pro,jah}\\ \hline
13 & $x_2\stackrel{k_{13}}{\longrightarrow}\phi$ & $Mdm2$ degradation & $k_{13} \langle x_2\rangle$ & $433\times 10^{-4}{sec}^{-1}$ & \cite{mol,pro,jah} \\ \hline
14 & $\phi\stackrel{k_{14}}{\longrightarrow}x_6$ & ROS formation & $k_{14}$ & $1.0\times 10^{-2}{sec}^{-1}$ & \cite{proc,pro} \\ \hline
15 & $x_6\stackrel{k_{15}}{\longrightarrow}\phi$ & Degradation of ROS & $k_{15} \langle x_6\rangle$ & $2\times 10^{-2}{sec}^{-1}$ & \cite{proc,pro}\\ \hline
16 & $x_6\stackrel{k_{16}}{\longrightarrow}x_7$ & Initiation of DNA damage & $k_{16} \langle x_6\rangle$ & $2\times 10^{-2}{sec}^{-1}$ & \cite{pro,proc}\\ \hline
17 & $x_7\stackrel{k_{17}}{\longrightarrow}\phi$ & DNA repair & $k_{17} \langle x_7\rangle$ & $2\times 10^{-5}{sec}^{-1}$ & \cite{pro,proc}\\ \hline
18 & $x_7\stackrel{k_{18}}{\longrightarrow}x_8+x_7$ & Activation of ARF & $k_{18} \langle x_7\rangle$ & $33\times 10^{-5}{sec}^{-1}$ & \cite{zha,pro,kha} \\ \hline
19 & $x_8\stackrel{k_{19}}{\longrightarrow}\phi$ & Degradation of ARF & $k_{19} \langle x_{8}\rangle$ & $1\times 10^{-4}{sec}^{-1}$ & \cite{zha,pro,kha} \\ \hline
20 & $x_8+x_2\stackrel{k_{20}}{\longrightarrow}x_9$ & $ARF\_Mdm2$ complex formation & $k_{20} \langle x_{8}\rangle \langle x_2\rangle $ & $1\times 10^{-2}{sec}^{-1}$ & \cite{zha,pro,kha}\\ \hline
21 & $x_9\stackrel{k_{21}}{\longrightarrow}x_8$ & Dissociation of $ARF\_Mdm2$ complex & $k_{21} \langle x_9\rangle$ & $1\times 10^{-3}{sec}^{-1}$ & \cite{zha,pro,kha} \\ \hline
22 & $x_{10}\stackrel{k_{22}}{\longrightarrow}\phi$ & Degradation of Micro RNA & $k_{22} \langle x_{10}\rangle $ & $5\times 10^{-2}{sec}^{-1}$ & \cite{lee1,win} \\ \hline
\end{tabular}
\end{center}
\end{table*}